\documentclass[twocolumn,showpacs,preprintnumbers,amsmath,amssymb]{revtex4}

\usepackage{graphicx}
\usepackage{dcolumn}
\usepackage{bm}

\begin{document}
\title{Optical velocimetry; LA-UR-04-6453}

\author{W. T. Buttler, S. K. Lamoreaux, F. G. Omenetto and J. R. Torgerson}


\affiliation{University of California,Los Alamos National
Laboratory,Physics Division P-23, M.S. H803, Los Alamos, NM 87545}

\date{July 14, 2004}

\begin{abstract}

This work considers current potential uses of laser Doppler
velocimetry. A discussion of other optical velocimetry techniques
is presented and compared with their practical application to
modern shock physics diagnostics, such as VISAR.

\end{abstract}
 \maketitle

\section{Introduction}

Doppler effect concepts were first introduced by mathematician
Christian Johann Doppler in the 1800s \cite{doppler}. Doppler
postulated that the frequency of sound, and light, depended on
whether the source was moving or stationary, and on the medium
through which the sound was carried. As a proof, in $1845$ Doppler
performed an experiment at a train station in which two identical
horns played the same note while one horn was on a moving train
and the other was stationary. This result is now generally
appreciated by most persons as most have heard the difference
between the sound of a car horn, siren, {\it etc}., that is
travelling toward or away from a stationary listener.

At a later date ($\sim1848$), Armand Fizeau independently
formalized Doppler's theory for electromagnetic. Fizeau was
unaware of Doppler's work \cite{doppler}, and his postulates dealt
with the red shift of light by moving stars. He postulated that
the speed of the star depended on its color, or frequency. Of
Course, Einstein fully established the relativistic Doppler effect
with his {\it Special Theory of Relativity} in the early 1905
\cite{einstein0}.

Presently the Doppler effect is used quite effectively, and
regularly, to determine the velocities of many different objects,
to answer many different physics questions. Police use the Doppler
effect, with optical- and radio-frequencies, to remotely determine
the velocity of cars. The military has used Doppler radar for
similar purposes. Doppler annemometers are used to measure
particle velocities in fluid flow. Laser Doppler velocimeters are
used to measure wind speeds \cite{buttler0}, {\it etc}.

Modern laser Doppler techniques were introduced in the early 1960s
by Yeh and Cummins \cite{Yeh}. Shortly thereafter, Forman, {\it et
al}, \cite{forman} proved Yeh's efforts and developed practical
applications of laser Doppler velocimetry as a fluid flow
diagnostic (see Fig. 1). Since this modern inception, Doppler
methods as a diagnostic tool have become popular in atmospheric
physics, eavesdropping, {\it etc}.

\begin{figure}[!hb]
\begin{center}
\includegraphics[width=3.25in ]{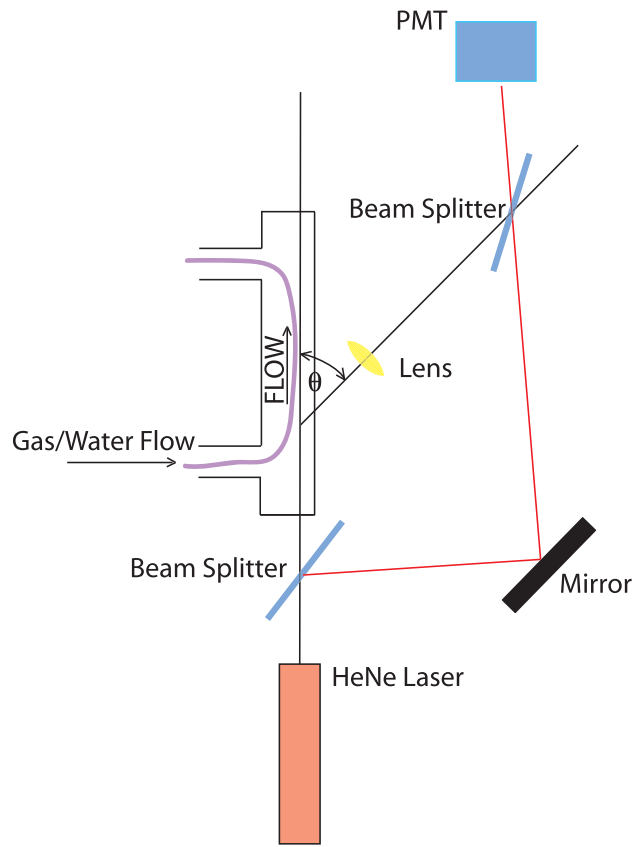}
\caption{This schematic shows the basic configuration used by
Forman \cite{forman}. Its salient characteristic is that it splits
off the reference beam and heterodynes it with the Doppler shifted
beam. The concept is simple, and the technique sensitive. (Not
shown are any attenuators that might be needed to balance the
reference beam with the reflected beam intensity; neither is the
beamsplitting ratio specified, nor any other detailed optics
specifications.)}
\end{center}
\end{figure}

\section{Optical Diagnostics: Shock physics applications}

Measurement problems in shock physics present their own special
difficulty in determining parameters important for modeling
continuous mechanical systems, such as steel or other metals,
under extreme pressures. At once Doppler methods looked promising
to measure the parameters important in equation-of-state studies,
and yet were difficult to practically implement. The largest
problems centered on available laser-light frequencies, and
detection-and-measurement technologies available at that time. For
example, consider the Doppler shift from He-Ne light ($\lambda =
633$ nm) scattered off of a surface shocked up to a velocity on
the order of $1,000 \text{ m-s}^{-1}$. This shift is represented
by a Doppler beat frequency of:
\begin{eqnarray}
   \omega_{D} & = & \frac{v}{c} \times \omega_0 \nonumber \\
   \therefore | f_{D} | & = & a \cdot \frac{v}{\lambda_0} = 1.58 \text{ GHz.}
\end{eqnarray}
(The constant, $a$, is determined by the angle of observation, and
whether the light is travelling along or against the flow; for
normal incidence reflection, $a = 2$.) Thus, one can see quite
clearly that in the early stages of Doppler development, that
measurement-and-detection technologies at the time limited Doppler
techniques to low speed measurements.

Not to be denied, scientists applied some neat optical techniques
to measure velocities of this order of magnitude; namely, VISAR
(velocity interferometer system for any reflector) \cite{barker}.
The success of VISAR relies on the Doppler shift to determine a
velocity from a reflector, but it does not actually measure the
Doppler shift. Rather, VISAR measures the difference in the
Doppler shift between two relatively Doppler shifted, reflected
light beams. Thus, if you will, VISAR measures the acceleration of
a surface as a function of time.

The VISAR concept, essentially, is based on an unbalanced
interferometer, i.e., one arm of the interferometer is longer than
the other (see Fig. $2$). The effect is to interfere early-time
reflected light with late-time reflected light. In its simplest
concept, detector amplitude measurements are made in quadrature;
the amplitude measurements then represent the sine and cosine of
the detected amplitude at the last beamsplitter. In this concept,
the sine- or-cosine is plotted versus the cosine- or-sine of the
measured amplitude--the phase angle between the early- and
late-time reflected light. To complete the system, polarizers and
waveplates are used to ascertain whether the surface is
accelerating in a positive or negative sense. The velocity is
determined by integrating (counting) the number of fringes that
{\it spin} by. The longer the signal is integrated in time, the
more uncertain the result.

\begin{figure}[!hb]
\begin{center}
\includegraphics[width=3.25in]{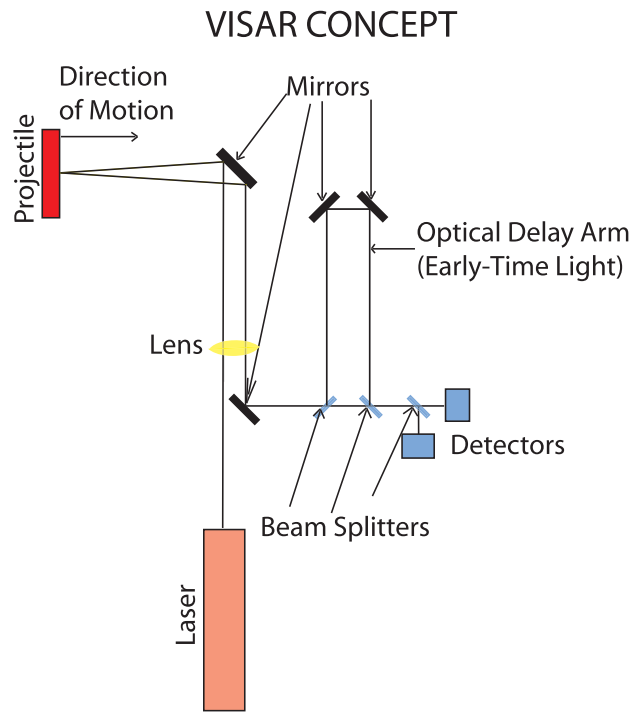}
\caption{\label{VISAR} This figure \cite{barker} schematically
describes the VISAR concept. Early-time reflected light [light
that travels the upper (long) path] is ''interfered'' with
late-time reflected light [light that travels the lower (short)
path]. The two detectors define the quadrature measurement. [It is
worth noting that many of the subsequent improvements to this
earlier system are left out of the schematic. The improvements
include a quality block of glass (this changes the rate of modal
dispersion), and polarizers and waveplates, to name a few of these
important improvements.]}
\end{center}
\end{figure}

VISAR was an elegant solution to an intractable
measurement-and-detection bandwidth problem in the early 1970s.
However, it has also been regularly applied to Asay foil
measurements \cite{asay} when in fact direct Doppler measurements
are more appropriate as the foil velocity in such systems is
seldom greater than a few tens of meters per second (a beat
frequency on the order of tens of mega-Hertz). It would be much
simpler and more cost effective to simply apply Doppler techniques
to the Asay foil diagnostic.

Another technique currently in use to measure the free-surface
velocity of a projectile, or shocked system of particles, is
referred to as Fabry-Perot \cite{fabry-perot}. Basically, this
concept involves directly measuring the change in the wavelength
of the reflected light. This is accomplished with a high quality
etalon. When properly aligned, the fringe pattern formed when
light passes through the etalon forms an Airy pattern. This Airy
pattern (a.k.a. Fraunhofer diffraction pattern of a circular
aperture) is imaged onto a slit preceding a streak camera that is
used for detection. As the surface is shocked, or as the
projectile moves, the Airy pattern fringe spacing changes. These
changes in the fringe patterns are recorded on the streak camera
and relate the velocity of the free-surface similarly as the
direct Doppler measurements.

\section{Discussion}

We have discussed several methods whereby it is possible to
measure the velocity of a free surface, such as a projectile or
shocked surface. The techniques commonly in use today at national
laboratories in the United States and Europe include VISAR and
Fabry-Perot, and the most common technique in use is VISAR
\cite{barker}. A reasonable question to ask, is ``why consider
different techniques today as we have at least two proven
diagnostic methods to accomplish this task?''. To answer this
question, it is appropriate to ask the following: ``what method
would we consider today for these measurements if these types of
measurements had never been accomplished before, and we were not
limited by our technology?''. The answer for velocities on the
order of $u(t) \lesssim 1,000$ m-s$^{-1}$, is that we would
consider Doppler.

Doppler measurements give the instantaneous velocity, so long as
certain parameters are known. For example, is the object moving
toward or away from the measurement apparatus? Is the angle of
incidence normal to the surface or is the angle known?

The Doppler measurement gives the velocity in the absence of {\it
a priori} information. For example, if the VISAR system is not
activated prior to the initial motion of the surface, the all
important initial phase angle information is lost and it cannot be
recovered. The initial motion of interest must be covered prior to
being able to determine a velocity; failure to measure this angle
is equivalent to not knowing (measuring) the integration constant.
In addition, due to bandwidth limitations, VISAR cannot accurately
determine the jump-off velocity as the measurement is sensitive to
small differences in the Doppler shift.

Fabry-Perot requires similar information as VISAR. For example,
the streak-camera timing must be precise enough to make certain
that the recording system is active prior to the initial surface
motion, otherwise the motion of the fringes from the static
position will not be known, and neither will the velocity.

It is simple to argue that Doppler has other limitations as well.
It is clear, however, that it is better to start the recording of
the diagnostic prior to projectile motion, or free-surface motion,
otherwise it would not be able to specify the position as a
function of time, at least not precisely ($x_0$ is needed, for the
same reasons that the initial phase angle is needed for VISAR, and
the initial fringe position is needed for Fabry-Perot: position
follows from integrating and the initial constant is required for
a quantitative result).

Because VISAR ``velocities'' represent an integration of the
acceleration measurement, it is clear that if each fringe that is
integrated is known to within some constant uncertainty, then the
uncertainty in velocity increases with time in a non-linear
fashion. If velocity is then integrated again, to generate
position as a function of time, then those uncertainties again
increase in a non-linear fashion. An advantage of an optical
Doppler measurement is that it measures velocity directly, and its
uncertainty is, for the most part, fixed with each measurement,
i.e, the uncertainty is related to how well one can determine the
distance between adjacent peaks in the beat frequency, $f_D$. This
is true for each measurement of the beat frequency. The positional
uncertainty, of course, increases with time, non-linearly.

\section{What makes sense?}

As noted earlier, a normally reflected Doppler measurement with a
He-Ne wavelength of $633$ nm requires a detector and recording
bandwidth of $\gtrsim 3.2$ GHz. While this is achievable today
with current technology, it would nevertheless remain expensive as
the recording-and-detection technologies are expensive; the laser
technology is within reach (for example, one might double and
Nd:YAG laser to $532$ nm). However, these types of bandwidths
remain impractical for practical applications (one might imagine
that velocities of interest may be one or two times $1,000$
m-s$^{-1}$).

The advent of high-speed optical communications technologies
provides a path forward for {\it direct Doppler techniques}. For
example, if the standard communications wavelength of $\lambda
\approx 1,550$ nm is used, then a bandwidth of $\approx 1.3$ GHz
is needed for a $1,000$ m-s$^{-1}$ velocity determination. This
leaves room for much higher velocities.

It should also be noted that Doppler will cost much less to field
per data set. It can be achieved without laser systems that cost
several hundred thousands of dollars, and should not require one
or two persons several months to set up, stabilize, and maintain
until the measurements are complete. There is no reason that laser
Doppler velocimetry cannot be plug-and-play.

The applications include high-velocity measurements at $1,550$ nm
(single-mode light, erbium doped fiber amplifiers will work well),
and Asay foil type measurements with a He-Ne type laser.

Other potentially practical uses include laser Doppler vibrometry
\cite{kurt} to measure strain coefficients in piezoelectric
materials. This technique could possibly be used to directly
measure the sensitivity of piezoelectric probes fielded as a
low-profile companion diagnostic to the Asay foil.

Our current effort includes development of a laser Doppler
velocimeter at $1,550$ nm. All optics have been purchased and
measurements on the order of mm-s$^{-1}$ velocities have been
accomplished; velocities on the order of cm-s$^{-1}$ have also
been accomplished with a ``fast'' solenoid. The crude apparatus
has been used and initial plans are to test the diagnostic at the
Proton-Radiography facility powder gun, and other LANL firing
sites; fielding the diagnostic at the Bechtel, Santa Barbara,
boom-box has also been considered.

Our objective is to develop an alternative optical diagnostic to
VISAR techniques where VISAR techniques are not required for
success. These situations include velocities up to about $3$
km-s$^{-1}$, with a $\lambda \approx 1,550$ nm laser tool, and a
visible He-Ne system for use with Asay foils. The laser
technologies are well established and the
measurement-and-detection technologies are available.

\section{Conclusion}

We have presented a discussion of current, laser remote velocity
determining systems. These systems include VISAR, Fabry-Perot and
Doppler. Direct laser Doppler techniques have several advantages
over both Fabry-Perot and VISAR. Namely, Doppler techniques are
more sensitive than VISAR or Fabry-Perot; Doppler does not suffer
from the same visibility issues as either VISAR or Fabry-Perot;
Doppler is easier to field with off the shelf components and does
not have the expense, nor should it require the continual support
required by VISAR or Fabry-Perot.



\end{document}